\documentclass[aps,nofootinbib, prd, amsmath, floats, floatfix, twocolumn,superscriptaddress, showpacs,reprint]{revtex4-1}

\usepackage{graphicx}
\usepackage{amsmath,amssymb}
\usepackage{amsfonts}
\usepackage{xspace} 
\usepackage[usenames]{color}
\usepackage{dcolumn}
\usepackage{bm}
\usepackage{mathrsfs}
\usepackage[colorlinks=true]{hyperref}
\usepackage[all]{hypcap} 
\usepackage[utf8]{inputenc}

\def\be{\begin{equation}}
\def\ee{\end{equation}}
\def\beq{\begin{eqnarray}}
\def\eeq{\end{eqnarray}}

\newcommand{\tn}{\textnormal}

\begin{document}

\title{Parameter estimation of gravitational wave echoes from exotic compact objects}

\author{Andrea Maselli}\email{andrea.maselli@uni-tuebingen.de}
\affiliation{Theoretical Astrophysics, Eberhard Karls University of Tuebingen, Tuebingen 72076, Germany}
\author{Sebastian H. V{\"o}lkel}\email{sebastian.voelkel@uni-tuebingen.de}
\affiliation{Theoretical Astrophysics, Eberhard Karls University of Tuebingen, Tuebingen 72076, Germany}
\author {Kostas D. Kokkotas}
\affiliation{Theoretical Astrophysics, Eberhard Karls University of Tuebingen, Tuebingen 72076, Germany}


\date{\today}
\begin{abstract}
Relativistic ultracompact objects without an event horizon may be able to form in nature and merge as binary systems, mimicking the coalescence of ordinary 
black holes. The postmerger phase of such processes presents characteristic signatures, which appear as repeated pulses within the emitted gravitational 
waveform,  i.e., echoes with variable amplitudes and frequencies. Future detections of these signals can shed new light on the existence of horizonless geometries 
and provide new information on the nature of gravity in a genuine strong-field regime. In this work we analyze phenomenological templates used to characterize 
echolike structures produced by exotic compact objects, and we investigate for the first time the ability of current and future interferometers to constrain their 
parameters. Using different models with an increasing level of accuracy, we determine the features that can be measured with the largest precision, and we span the 
parameter space to find the most favorable configurations to be detected. Our analysis shows that current detectors may already be able to extract all the 
parameters of the echoes with good accuracy, and that multiple interferometers can measure frequencies and damping factors of the signals at the level of percent. 
\end{abstract}

\maketitle 

%%%%%%%%%%%%%%%%%%%%%%%%%%%%%%%%%%%%%%%%%%%%%%%%%%%%%%%
\section{Introduction}
%%%%%%%%%%%%%%%%%%%%%%%%%%%%%%%%%%%%%%%%%%%%%%%%%%%%%%%

Gravitational wave (GW) astronomy is nowadays emerging as a new observational window, able to provide fundamental 
insights on some of the most energetic phenomena of our Universe. The amount of incoming data produced 
by ground based interferometers also promises to address questions of fundamental physics with unprecedented accuracy. 
Among all the possible compact sources, black holes (BH) are probably the most extreme physical systems, whose 
existence has been definitively assessed by the recent LIGO discoveries \cite{LIGO1,PhysRevLett.116.241103,PhysRevLett.118.221101}. 
These detections mark the dawn of BH spectroscopy and at the same time represent the first genuine strong-field tests of general relativity \cite{LIGO2}.

However, some crucial questions regarding the fundamental nature of BHs still remain to be addressed \cite{Cardoso:2017njb}. 
As an example, theoretical models predicting the existence of exotic compact objects (ECOs) whose compactness approaches the BH limit 
have not been completely ruled out. Such bodies may form in nature as binary systems and merge due to GW emission. 
During the coalescence the ECOs leave distinct signatures within the inspiral part of the signal, which has already proved to be extremely effective 
in discriminating between regular BHs and exotic scenarios \cite{Cardoso:2017cfl,Maselli:2017cmm}.

After the merger, horizonless compact objects will emit gravitational radiation until they reach a quiet and stationary state. During this 
process, multiple {\it trapped $w$ modes} may be excited, which would be visible within the GW signal by the appearance 
of echolike structures, i.e., repeated pulses with characteristic frequencies and amplitudes, which differ from the BH quasinormal modes (QNM)
spectrum \cite{Kokkotas:1995av,PhysRevD.60.024004,2000PhRvD..62j7504F,PhysRevD.63.064018,PhysRevD.63.064018}. 

Historically, the idea that QNM may represent a powerful tool to distinguish between ultracompact stars and regular BHs (or less compact bodies) traces back its origin in some 
seminal works of the early 1990s \cite{1991RSPSA.434..449C,1991RSPSA.432..247C,1992MNRAS.255..119K,1994MNRAS.268.1015K,1995RSPSA.451..341K,1996ApJ...462..855A}.
A revised application of this approach has  recently been applied to interpret the LIGO data in terms  of new physics at the level of the BH horizon.
This work has  drawn a lot of attention \cite{2016arXiv161200266A,2017arXiv170103485A} and triggered new exciting research efforts in the field 
\cite{Barcelo2017,2017arXiv170204833P,paper1,2017arXiv170303696M,2017arXiv170405789B,2017arXiv170405856H,PhysRevD.95.084034,paper2}  (see also \cite{2016PhRvD..94h4016C,2016arXiv161205625A} for some criticism on the same topic). 
Future detections with a higher signal-to-noise ratio, and the final completion of 
multiple GW detectors, like VIRGO \cite{virgo} and KAGRA  \cite{kagra}, will provide more accurate data,  possibly leading to assess whether 
horizonless compact bodies may exist in astrophysical environments \cite{PhysRevLett.117.101102,2016CQGra..33q4001C,PhysRevLett.118.161101,PhysRevD.95.104026}.

Several efforts have already been  devoted to characterize the GW emission of exotic objects out of equilibrium \cite{2007PhRvD..76b4016D,2007CQGra..24.4191C,2016PhRvL.116q1101C,Konoplya:2016hmd,2016PhRvD..94h4031C,2016arXiv161200266A}. 
If additional structures appear within the spectrum, our ability to extract the signatures  which deviates from the standard BH picture, will strongly 
depend on the availability of realistic templates to be used in data searches. In this sense, the recent works by \cite{Nakano:2017fvh,2017arXiv170606155M} 
provide the first systematic attempts to construct fully reliable templates to identify the echoes.

Motivated by these results, in this paper we explore for the first time the detectability of GW signals emitted by ECOs 
formed after binary coalescences. 
We consider different phenomenological templates, which are physically motivated by the analysis of perturbed ultracompact stars and from a series of recent work on the subject 
\cite{2007PhRvD..76b4016D,2007CQGra..24.4191C,2016PhRvL.116q1101C,2016PhRvD..94h4031C,2016arXiv161200266A,Nakano:2017fvh,2017arXiv170606155M}. 
The scope of this study is twofold: (i) determine the errors on the waveform's parameters, 
which would be measured by current and future GW interferometers, and (ii) investigate the dependence of such detections 
by the echo's parameters. Although the models employed suffer from some limitations, a complete and 
fully accurate description of the GW signal is nowadays not available. Nevertheless, the analysis developed in this work 
captures important features of the overall phenomena. Our results suggest that Advanced LIGO at design sensitivity would already be 
able to constrain the parameters of the echoes with good accuracy, possibly leading to infer new information on the nature of the perturbed 
compact object.

This paper is organized as follows. In Sec.~\ref{Sec:templates} we define the analytical templates used to model 
the echoes, which will be used to determine the parameters' detectability. In Sec.~\ref{Sec:analysis} we briefly describe 
the data-analysis procedure employed, while in Sec.~\ref{Sec:results} we present our numerical results, analyzing the 
errors on the gravitational waveforms for different interferometers. In Sec.~\ref{Sec:conc} we summarize our conclusions. 
Throughout the paper we will use geometrical units $(G=c=1)$.

%%%%%%%%%%%%%%%%%%%%%%%%%%%%%%%%%%%%%%%%%%%%%%%%%%%%%%%
\section{The echo templates}\label{Sec:templates}
%%%%%%%%%%%%%%%%%%%%%%%%%%%%%%%%%%%%%%%%%%%%%%%%%%%%%%%

\begin{figure}[ht]
\includegraphics[width=7cm]{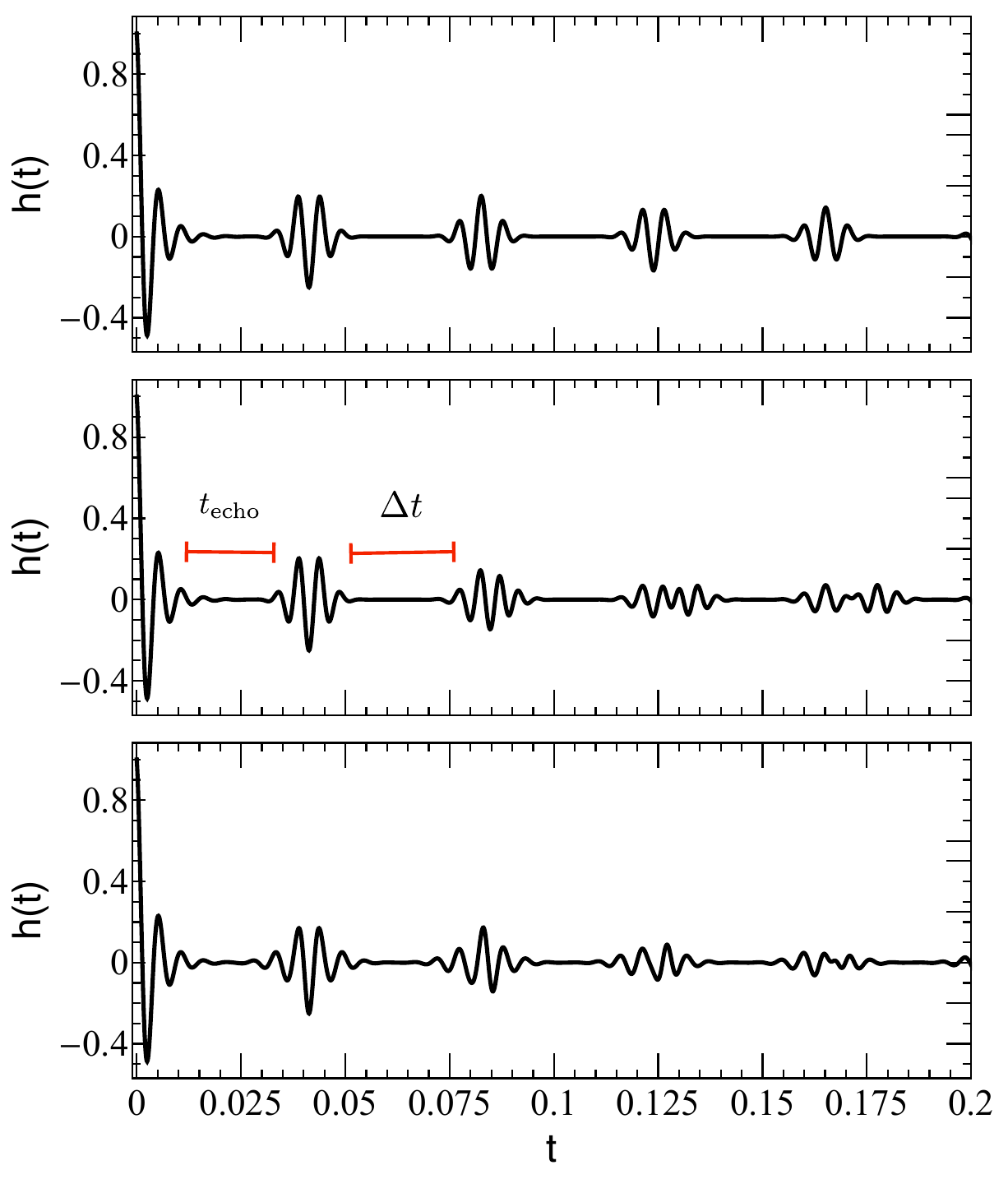}
\label{fig:templatess} 
\caption{Sketch of some phenomenological waveforms used in this work. 
The center panel also shows the meaning of the two parameters 
$t_\tn{echo}$ and $\Delta t$, which identify the time shift between different pulses 
in the template.}
\end{figure}

In this section we shall describe the GW templates used to estimate the errors on the echo's 
parameters. It is worth mentioning that some efforts have recently been made in 
\cite{2016arXiv161200266A,Nakano:2017fvh,2017arXiv170606155M} to propose analytical models that 
characterize the late time waveform of perturbed exotic objects. In this direction, the main purpose of our paper is to investigate 
how pure phenomenological waveforms may constrain the fundamental features of the pulses produced after the 
merger by ultracompact objects with a reflecting surface. We develop our analysis in a pedagogical way, starting 
from the simplest model, up to more sophisticated waveforms that may eventually mimic the {\it true} GW 
emission by a real ECO. For more details on the physics of the echolike structure we refer the reader to the 
literature that is mentioned in the Introduction. All our models are described by an early ringdown, which 
represents the fundamental BH QNM damped oscillation, followed by a series of repeated echoes. Hereafter 
we consider three different templates, defined as follows:
\begin{itemize}
\item
\texttt{echoI}: the waveform is given analytically by $h_\texttt{I}(t)=h_\tn{QNM}(t)+h_\tn{I}(t)$, where 
\begin{equation}
h_\tn{QNM}(t)=\bar{{\cal A}}e^{- t/\bar{\tau}}\cos(2\pi \bar{f}t+\bar{\phi})\label{ecoI2}\ ,\ 
\end{equation}
corresponds to the BH QNM-like oscillation, specified by amplitude, frequency, phase, and damping time 
$(\bar{{\cal A}}, \bar{f},\bar{\phi},\bar{\tau})$, while
\begin{equation}
h_\tn{I}(t)=\sum_{n=0}^{N-1}(-1)^{n+1}{\cal A}_{n+1}e^{-\frac{y^2_{n}}{2\beta_1^2}}\cos(2\pi f_1 y_n)\ ,\label{heco1}
\end{equation}
describes the $N$ echoes after the first mode. Note that in this case we assume the same frequency and shape $(f_1,\beta_1)$ for 
each pulse, but different values of the amplitude ${\cal A}_{n+1}={\cal A}_{1,\ldots,N}$. For the sake of simplicity we have 
chosen the modulating function as a Gaussian profile, with variance given by $\beta_1$. This setup is also in agreement with the 
analysis developed in \cite{Kokkotas:1995av} to investigate GW signals produced by  ultracompact stars perturbed by Gaussian pulses. 
In the former equation we have also introduced the auxiliary variable 
$y_{n}(t)=(t-t_\tn{echo}-n\Delta t)$, where $t_\tn{echo}$ is the time shift between the first mode and the first echo, while $\Delta t$ identifies 
the time delay between the successive $N-1$ echoes (see Fig.~\ref{fig:templatess}). 
\item
\texttt{echoIIa}:
the first oscillation corresponds again to the BH ringdown mode as in \texttt{echoI}, although the echo sector is 
improved by introducing a second frequency $f_2$. The latter takes into account that the frequencies of the pulses we observe 
in the spectrum are related to the trapped modes of the system, which consist in general of multiple components. The template is then given by
\begin{equation}
h_\texttt{II}=h_\tn{QNM}(t)+h_\tn{IIa}(t,f_1,f_2,\beta_1)\ ,
\end{equation}
where 
\begin{align}
h_\tn{IIa}(t)=\frac{1}{2}\sum_{n=0}^{N-1}(-1)^{n+1}{\cal A}_{n+1}&e^{-\frac{y^2_{n}}{2\beta_1^2}}\big[\cos(2\pi f_1 y_n)\nonumber\\
&+\cos(2\pi f_2 y_n+\phi)\big]\ ,\label{heco2a}
\end{align}
and the phase $\phi$ determines an offset between the two terms for $t=0$.
Equation \eqref{heco2a} describes a beatlike structure, which should mimic as a first approximation 
the interference of the trapped modes.
\item
\texttt{echoIIb}:
this further generalizes the previous approaches by adding a different Gaussian function for the second mode of 
the echoes, i.e, $h_\texttt{II}=h_\tn{QNM}(t)+h_\tn{IIb}(t,f_1,f_2,\beta_1,\beta_2)$, with
\begin{align}
h_\tn{IIb}(t)=\frac{1}{2}\sum_{n=0}^{N-1}(-1)^{n+1}&{\cal A}_{n+1}\big[e^{-\frac{y^2_{n}}{2\beta_1^2}}\cos(2\pi f_1 y_n)\nonumber\\
+&e^{-\frac{y^2_{n}}{2\beta_2^2}}\cos(2\pi f_2 y_n+\phi)\big]\ .\label{heco2b}
\end{align}
\end{itemize} 
For all the waveforms we will further assume that amplitudes ${\cal A}_{1\cdots N}$ carry a fraction of the QNM component $\bar{\cal A}$. 
This choice is physically motivated by numerical results obtained from an updated version of a code for ultracompact constant density 
stars presented in \cite{Kokkotas:1995av}, in which the ratio between the QNM mode and the first pulse is roughly equal 
to $\bar{{\cal A}}/{\cal A}_1\sim1/4$, and then decreases as $\bar{{\cal A}}/{\cal A}_N\sim\frac{1}{4+N}$ for the following $N$ 
echoes. This assumption also reduces the number of independent amplitudes to the overall BH-like factor $\bar{{\cal A}}$.

\par
The generalization of the previous templates to more sophisticated models is straightforward and could 
include the following features: (i) add different frequencies and their interference to characterize each echo;
(ii) include the damping factor of each frequency (although we expect they would play a subordinate role 
within the data analysis of the waveform); and (iii) introduce different functions to model the shape of the echoes, 
instead of the Gaussian profile used in this paper. Such improvements would lead one to consider a more realistic scenario, 
which would ultimately depend on the nature of the ECO's perturbation.
These extensions will provide a more detailed picture of the physical mechanism producing the echoes and are under investigation. However, we believe that the GW templates presented in this 
section are already able to capture the most relevant features of the real process. 

%%%%%%%%%%%%%%%%%%%%%%%%%%%%%%%%%%%%%%%%%%%%%%%%%%%%%%%
\section{Data analysis procedure}\label{Sec:analysis}
%%%%%%%%%%%%%%%%%%%%%%%%%%%%%%%%%%%%%%%%%%%%%%%%%%%%%%%

To compute the errors on the parameters of the echo's template, we use a Fisher matrix approach 
\cite{1993PhRvD..48.3451K,1994CQGra..11.1901K,Vallisneri:2007ev}. 
In the limit of a large signal-to-noise ratio (SNR), the probability distribution of the parameters $\vec{\theta}$ for 
a given set of data $d$ can be expanded around the true values $\vec{\theta}_\tn{v}$ as
\begin{align}
p(\vec{\theta}\vert d)&=p_0(\vec{\theta}){\cal L}(d\vert\vec{\theta})=p_0(\vec{\theta})e^{-\frac{1}{2}(\tilde{h}(\vec{\theta})-d\vert \tilde{h}(\vec{\theta})-d)}\nonumber\\
&\approx p_0(\vec{\theta})e^{-\frac{1}{2}\Gamma_{\alpha\beta}\Delta \theta^\alpha\Delta \theta^\beta}\ ,
\end{align}
with $p_0(\vec{\theta})$ being the prior probability on $\vec{\theta}$, and $\Delta \theta^\alpha=\theta^\alpha-
\theta^\alpha_\tn{v}$. The Fisher information matrix $\Gamma_{\alpha\beta}$, which characterizes the curvature of 
the likelihood function ${\cal L}(d\vert\vec{\theta})$, is expressed in terms of the partial 
derivatives of the GW template with respect to the echo parameters,
\begin{equation}
\Gamma_{\alpha\beta}=\left(\frac{\partial \tilde{h}}{\partial \theta^\alpha}\bigg\vert 
\frac{\partial \tilde{h}}{\partial \theta^\beta}\right)_{\vec{\theta}=\vec{\theta}_\tn{v}}\ ,\label{fisher}
\end{equation}
where $(\tilde{a}\vert \tilde{c})$ defines the scalar product on the waveform's space,
\begin{equation}
(\tilde{a}\vert \tilde{c})=2\int_{0}^{\infty}\frac{\tilde{a}(f)\tilde{c}^\star(f)
+\tilde{a}^\star(f)\tilde{c}(f)}{S_n(f)}df\ ,\label{scalarprod}
\end{equation}
$S_n(f)$ is the noise spectral density of the chosen detector, and $\tilde{h}(f)$ is the Fourier transform 
of the template in the frequency domain\footnote{We use the following normalization for the Fourier transform of 
the templates:
$$
\tilde{h}(f)=\int_{-\infty}^{\infty}h(t) dt\quad \ ,\quad h(t)=\frac{1}{2\pi}\int_{-\infty}^{\infty}\tilde{h}(f) df\ .
$$
All the waveforms considered yield a full analytical form of $\tilde{h}(f)$, which can be easily computed by means 
of symbolic manipulation softwares like \texttt{Mathematica}.}.  
The covariance matrix of the parameters is simply given by the 
inverse of the Fisher, i.e., $\Sigma_{\alpha\beta}=(\Gamma_{\alpha\beta})^{-1}$, whose diagonal and 
off-diagonal components correspond to the standard deviations and the correlation coefficients of $\vec{\theta}$, 
respectively. Note that, according to the Cramer-Rao bound, the uncertainties obtained through the 
Fisher matrix represent a lower constraint on the variance of any unbiased estimator of the parameters.
The scalar product~\eqref{scalarprod} also allows one to define the SNR of the specific signal, as 
\begin{equation}
\rho^2=(\tilde{h}\vert \tilde{h})=4\int_0^\infty \frac{\vert \tilde{h}(f)\vert^2}{S_n(f)}df\ .\label{SNR}
\end{equation}

In this paper we consider the detectability of echoes by current and future generations of detectors, i.e., Advanced 
LIGO with the \texttt{ZERO\_DET\_high\_P} anticipated 
design sensitivity curve \cite{zerodet},  the Einstein Telescope (ET) \cite{Hild:2009ns}, LIGO-Voyager (VY) 
\cite{LIGOWhite}, Advanced LIGO with squeezing (LIGO A+) \cite{Miller:2014kma}, and the Cosmic Explorer (CE) 
with a wide-band configuration \cite{0264-9381-34-4-044001}. 
In the following section we will quote our results on the specific parameter of the template $\theta^\alpha$ either 
in terms of the absolute error $\sigma_\alpha$ or its relative (percentage) value $\epsilon_\alpha=\sigma_\alpha/\theta^\alpha$.

%%%%%%%%%%%%%%%%%%%%%%%%%%%%%%%%%%%%%%%%%%%%%%%%%%%%%%%
\section{Constraints on the echo's parameters}\label{Sec:results}
%%%%%%%%%%%%%%%%%%%%%%%%%%%%%%%%%%%%%%%%%%%%%%%%%%%%%%%

In this section we present the results for the different templates of Sec.~\ref{Sec:templates}, obtained
by numerical integration of Eqs.~\eqref{fisher}-\eqref{SNR}. For all the models we choose the frequency and 
the damping factor of the QNM mode, as those of a nonrotating object with the same mass of the final BH 
formed in the GW150914 event \cite{TheLIGOScientific:2016src}, i.e. $M\simeq 65M_\odot$. This yields  $\bar{f}\simeq 186$ 
Hz and $\bar{\tau}\sim3.6\times 10^{-3}$s. Moreover, without loss of generality, we fix the phase of $\bar{h}(t)$ 
to $\bar{\phi}=0$, and the overall amplitude to a prototype value ${\cal A}=5\times 10^{-22}$, which roughly 
corresponds to a SNR of the QNM-like mode only (i.e., neglecting the contribution of the following echoes) of 
$\rho\sim 8$ with Advanced LIGO. This value is consistent with the best-fit parameters inferred from 
GW150914, O1 configuration \cite{TheLIGOScientific:2016src}. 
Note that, since $A$ represents a multiplicative factor of the total signal, our results can immediately be rescaled to 
any amplitude $A_\tn{new}$ as
\begin{equation}
\Gamma_{\alpha\beta}\rightarrow \frac{{\cal A}_\tn{new}}{5\times10^{-22}}\Gamma_{\alpha\beta}\quad \Rightarrow\quad
\sigma_\alpha\rightarrow\frac{5\times10^{-22}}{{\cal A}_\tn{new}}\sigma_\alpha\ ,
\end{equation}
and in the same way for the SNR,
\begin{equation}
\rho\rightarrow \frac{{\cal A}_\tn{new}}{5\times10^{-22}}\rho\ .
\end{equation}

After the first pulse, repeated echoes are also expected to occur with a time delay $\Delta t$, which depends 
on the features of the exotic object \cite{Cardoso:2017njb}, namely
\begin{equation}
\Delta t\sim 4M\vert \log\delta\vert\ ,\label{deltat}
\end{equation}
where $\delta\ll1$ represents the shift of the ECO's effective surface $r_0$ with respect to a nonrotating BH horizon 
located at $2M$ in the Schwarzschild coordinates\footnote{The coordinate distance is not gauge invariant, and therefore in general the specific value of $\delta$ is not uniquely defined. However the difference with respect to the proper distance is subordinate in our calculations due to the logarithmic dependence within $r_0$. Note also that the approximations for small $\delta$ are only valid for systems, where the reflecting surface is very close to the BH horizon in the Schwarzschild coordinate. Although constant density stars can feature a similar structure, the values of 
$\delta$ for such objects can never be small, due to the Buchdahl limit.}, i.e. 
$r_0=2M(1+\delta)$. From Eq.~\eqref{deltat}, we can approximate the potential well where echoes are reflected, 
with a box-potential specified by the coordinate width
\begin{equation}
x_\tn{c}\simeq 2 M\vert\log\delta\vert\ .
\end{equation}
Under this assumption, the correspondence between echoes and trapped modes inside the box allows one to express 
the gap between two consecutive modes $\Delta f$ with frequencies $f^\text{box}_{n+1}$ and $f^\text{box}_{n}$ as
\begin{equation}
\Delta f \equiv f^\text{box}_{n+1}-f^\text{box}_n \simeq \frac{1}{4 M\vert\log \delta\vert}\ .
\end{equation}
Then, having fixed the first frequency of each waveform to the corresponding BH QNM component, we can immediately 
derive the values of $f_1$ and $f_2$ used in the \texttt{echoI} and \texttt{echoIIa-b} templates:
\begin{equation}
f_1\simeq\bar{f}\quad\ ,\quad f_2\simeq\bar{f}-\Delta f\ .\label{freq}
\end{equation}
Note that $f_2 < f_1$. It is worthwhile to remark that these assumptions represent an approximation of the real physical scenario, in which we 
expect that $\bar{f}\neq f_1$, and $\Delta f $ takes a more complex form, which ultimately depends on the specific 
ECO considered. However, for the purpose of this paper, this will not change the outcome of the data-analysis 
procedure. Moreover, having fixed the object mass to $M=65M_\odot$, throughout this paper 
we consider three values of $\delta=(10^{-10},10^{-20},10^{-30})$, which roughly correspond to compact objects with 
surface corrections at Micron, Fermi, and Planckian levels, respectively \cite{Maselli:2017cmm}. 

Finally, the time between the QNM-like mode and the first echo, $t_\tn{echo}$, could be affected by nonlinearities due 
to the merger phase at the end of the coalescence \cite{Abedi:2016hgu}, i.e., $t_\tn{echo}\simeq\Delta t \pm \delta t$.
In the following, for each value of $\Delta t$ given by Eq.~\eqref{deltat}, we will consider different configurations 
by varying the coefficient $\delta t$ in order to have a maximum correction of the order $10\%$ on $t_\tn{echo}$. 

%%%%%%%%%%%%%%%%%%%%%%%%%%%%%%%
\begin{figure}[ht]
\centering
\includegraphics[width=8cm]{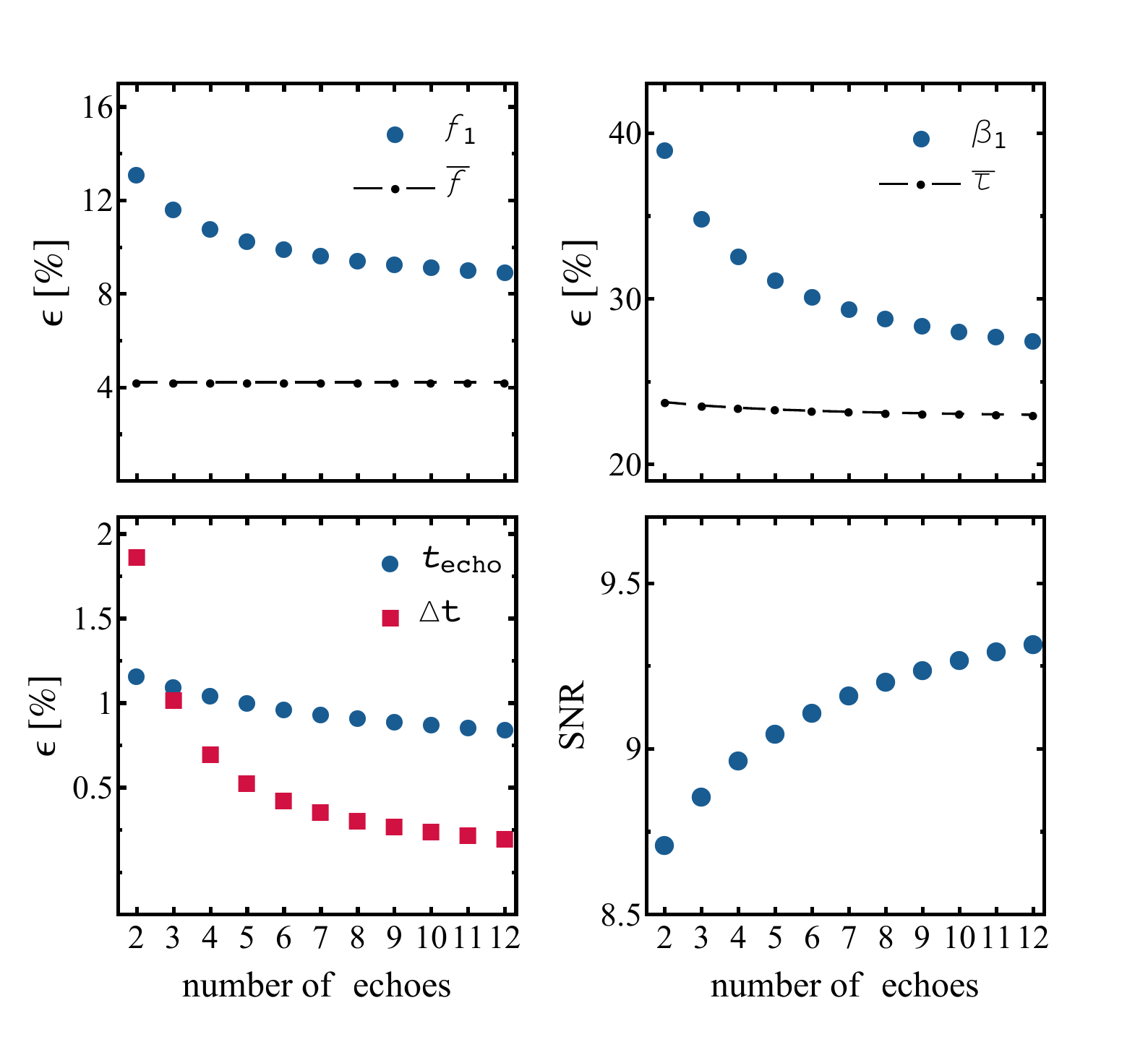}
\caption{Relative (percentage) errors on the parameters of the template \texttt{echoI} as a function of the 
number of echoes. All the results are derived for Advance LIGO, assuming $\delta=10^{-10}$ and 
$\beta_1=0.003$. The bottom right panel shows the change of the signal-to-noise ratio 
due to the increasing numbers of echoes.}
\label{fig:econumber}
\end{figure}
%%%%%%%%%%%%%%%%%%%%%%%%%%%%%%%%%%%%%%%%%%%%

Before assessing the detectability of each phenomenological template described in the previous section, it is 
instructive to analyze some basic features that are common to all the waveforms. Figure~\ref{fig:econumber} shows 
the relative errors $\epsilon_\alpha$ for the \texttt{echoI} model computed for LIGO, as a function of the number of 
echoes included within the template. In this particular case we assume $\beta_1=0.003$ and $\delta=10^{-10}$, which 
corresponds to $t_\tn{echo}=\Delta t\simeq2.95\times10^{-2}$s. 
From the first two panels we can immediately note that the uncertainty on frequency and damping time of the QNM 
component (black dots) is essentially unaffected by $N$, and it is therefore independent\footnote{The correlation 
coefficients derived from the Fisher matrix between $\bar{f}$ ($\bar{\tau})$ and the echo parameters are also very small 
for all the configurations.} from the template \eqref{heco1}. On the other hand, the errors on $(f_1,\beta_1)$ 
and on the delay times $(t_\tn{echo},\Delta t)$ reduce as far as the number of pulses grows in time. For the particular 
model analyzed here, the uncertainty on both $f_1$ and $\beta_1$ changes approximately $30\%$ between $N=2$ and 
$N=10$. Although these values seem to converge to the QNM mode value, this 
decrease saturates due to the progressive reduction of the echo's amplitudes.
This feature is also evident looking at the evolution of the overall SNR (right-bottom panel of Fig.~\ref{fig:econumber}), 
which reaches a nearly constant value of $\rho\sim 9.3$ after 12 pulses. This picture is nearly 
independent of the range of parameters used in this work and of the specific echo model adopted.

According to these results, we can safely consider gravitational waveforms that include 10 pulses 
after the QNM oscillation, since larger values of $N$ will not affect the analysis. This choice will also make 
our analysis more robust, since at later times some physical effects may not be captured by our models (as the 
interference of multiple trapped frequencies).

%%%%%%%%%%%%%%%%%%%%%%%%%%%%%%%%%%%%%%%%%%%%%%%%%%%%%%%
\subsection{echoI}\label{Sec:ecoI}
%%%%%%%%%%%%%%%%%%%%%%%%%%%%%%%%%%%%%%%%%%%%%%%%%%%%%%%

The simplest waveform \texttt{echoI} depends on the following set of parameters: $\vec{\theta}=\{\ln\bar{{\cal A}},\bar{f},\bar{\tau},\bar{\phi},f_1,\beta_1,
t_\tn{echo},\Delta t\}$, which lead to an $8\times8$ Fisher matrix. As described before, we fix the frequency and damping 
factor of the QNM, with $f_1$ and the time shift $\Delta t$ being specified by Eqs.~\eqref{deltat} and \eqref{freq}.
However, to explore the space of the parameter's configurations, we vary the shape factor $\beta_1$ and 
$t_\tn{echo}=\Delta t +\delta t$. This will allow one to determine the more (or less) favorable signals to be detected by GW 
interferometers. The width of the echo's Gaussian function represents the coefficient that 
dominantly affects the shape of the waveform and therefore leads to major changes in the parameter estimation. 
Moreover, we will only discuss the features of the post-QNM modes, since the errors on $\bar{f}$ and $\bar{\tau}$ do not vary significantly within 
all the configurations, peaking around $\epsilon_{\bar{f}}\sim4\%$ and $\epsilon_{\bar{\tau}}\sim22\%$-$23\%$, respectively.

%%%%%%%%%%%%%%%%%%%%%%%%%%%%%%%
\begin{figure}[ht]
\centering
\includegraphics[width=4.2cm]{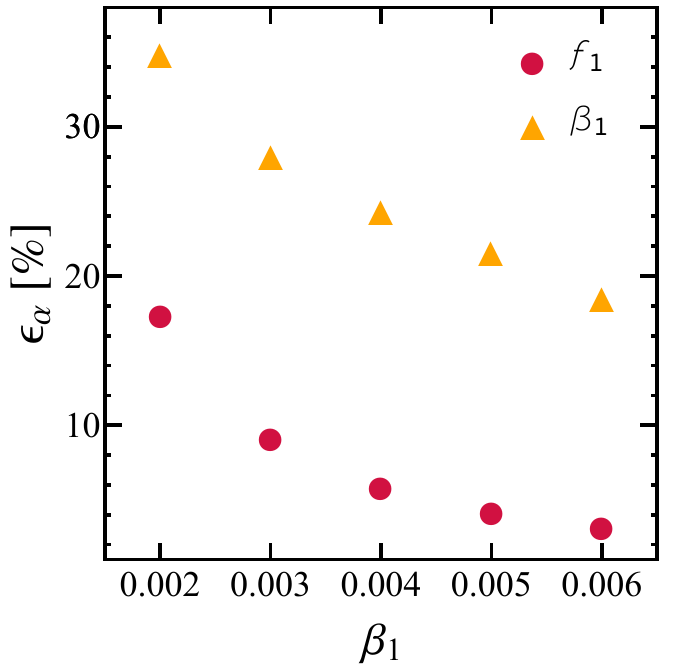}
\includegraphics[width=4.2cm]{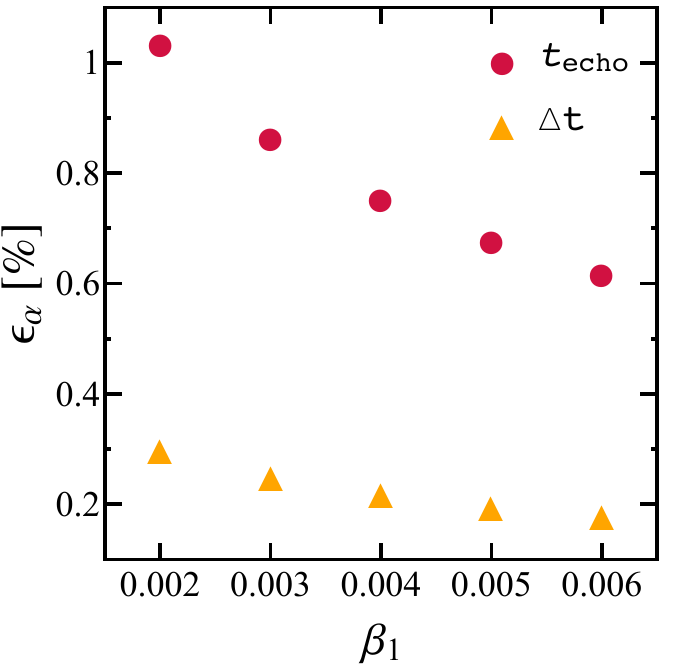}
\caption{Relative (percentage) errors on the parameters of the template \texttt{echoI} computed for Advanced 
LIGO, as a function of the width of the Gaussian width $\beta_1$. Both panels refer to  $t_\tn{echo}\approx\Delta t\approx2.95\times10^{-2}$ 
($\delta=10^{-10}$).}
\label{fig:errors1}
\end{figure}
%%%%%%%%%%%%%%%%%%%%%%%%%%%%%%%%%%%%%%%%%%%%

Figure~\ref{fig:errors1} shows the uncertainties of the \texttt{echoI} parameters as a function of  $\beta_1$ 
computed for Advanced LIGO, for a specific configuration with $t_\tn{echo}\approx\Delta t$ and $\delta=10^{-10}$. 
We immediately see from both panels that all relative errors rapidly decrease as the shape factor grows, 
with variations $\gtrsim40\%$ for $\epsilon_{f_1}$ and $\epsilon_{\beta_1}$. Note that the SNR changes between $\rho\sim 8.9$ for $\beta_1=0.002$ to 
$\rho \sim 10.3$ for $\beta\sim0.006$ with an overall increase of 
$15\%$. It is  important to remark that, although these differences do exist between the various configurations, 
all the modes considered yield errors smaller than a $1$-$\sigma$ upper bound with $\epsilon_\alpha=1$. This is 
particularly promising for the measurements of the time shift parameters (right panel), which can be constrained 
with an accuracy better than $1\%$.
 
\begin{figure}[ht]
\centering
\includegraphics[width=8.5cm]{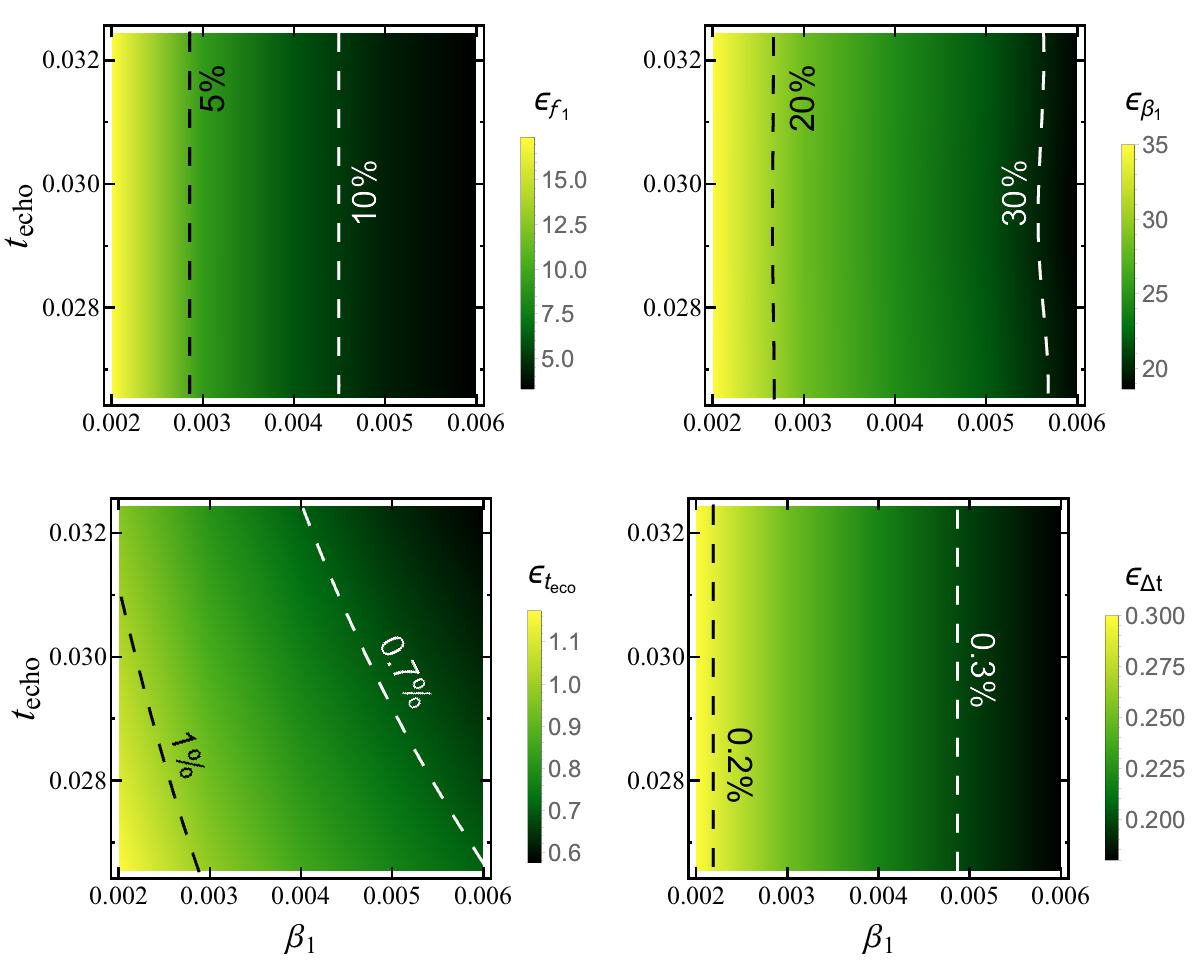}
\caption{Contour plots in the $\beta_1-t_\tn{echo}$ parameter space for the relative errors on the \texttt{echoI} parameters. 
White and black dashed curves represent configurations with fixed accuracy. The data refer to Advanced LIGO at design 
sensitivity.}
\label{fig:contours}
\end{figure}

The dependence of $\sigma_\alpha$ with respect to $t_\tn{echo}$ [which we vary in our data set as 
$t_\tn{echo}=\Delta t(1\pm0.1)$] is much milder and leads to nearly constant errors for all the parameters of the 
template. This can be appreciated from the contour plots of Fig.~\ref{fig:contours}, in which curves of fixed 
accuracy for $f_1,\beta_1$, and $\Delta t$ are given by vertical straight lines. 
Note that the {\it relative} errors on $t_\tn{echo}$ (bottom left) change less than $10\%$ within the parameter space considered, even though the {\it absolute} 
error remains practically constant.

%%%%%%%%%%%%%%%%%%%%%%%%%%%%%%%
%%%%%%%%%%%%%%%%%%%%%%%%%%%%%%%%%%%%%%%%%%%%

\begin{figure}[ht]
\centering
\includegraphics[width=4.2cm]{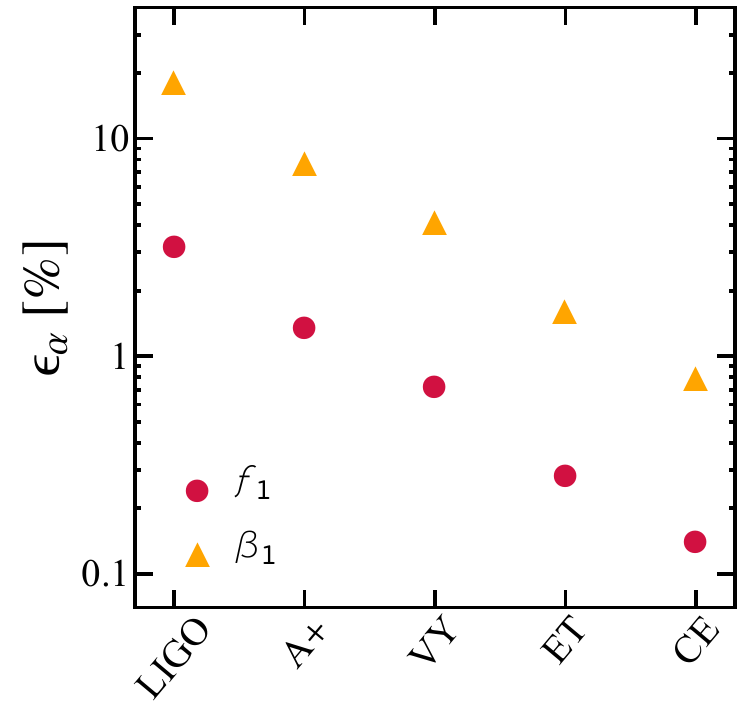}
\includegraphics[width=4.2cm]{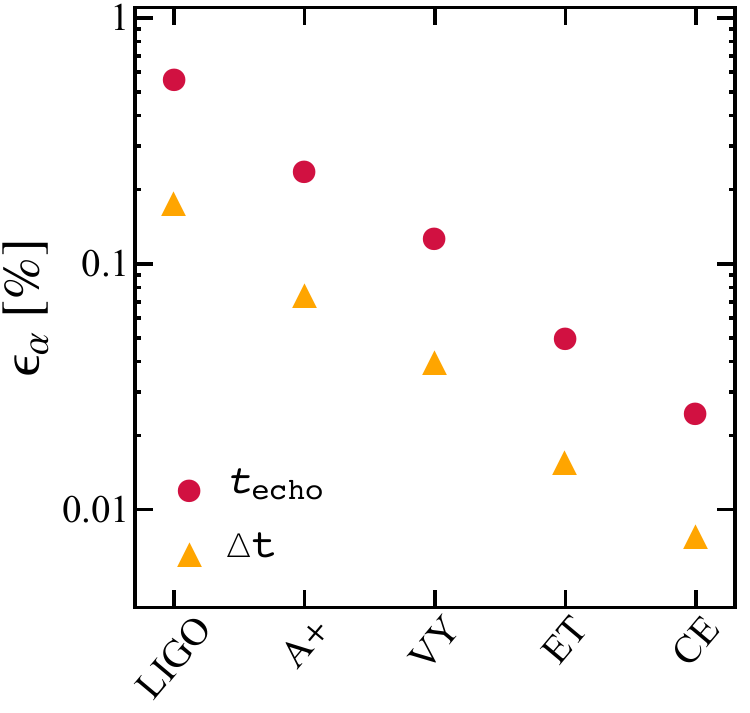}
\caption{Errors on the \texttt{echoI} template for different GW interferometers. The data refer to the best case scenario 
with $\beta_{1}=0.006$ (and different values of $t_\tn{echo}$).}
\label{fig:errors2}
\end{figure}

Although Advanced LIGO (at design sensitivity) seems already able to set narrow bounds on some of the 
echo's features, it is interesting to investigate how these results improve as far as we consider next generation detectors. 
This is shown in Fig.~\ref{fig:errors2}, in which we draw the relative errors of \texttt{echoI} for different interferometers. 
All the results correspond to the best-case scenario, i.e., for $\beta_1=0.006$. Note also that in general, for fixed $\beta_1$, 
the best measurements for each detector will correspond to a different value of $t_\tn{echo}$ (although changing this 
variable does not yield significant variations). 
Looking at the top panel we note that the errors of the echo's shape factor decrease to values of the order of $\leq 1\%$ already with 
LIGO A+, while for the frequency $f_1$, the same level of accuracy would require at least the ET. As expected, 
the recently proposed CE would lead to detect GW signals with exquisite precision, with errors being more than an 
order of magnitude smaller than values obtained by the current generation of detectors.

%%%%%%%%%%%%%%%%%%%%%%%%%%%%%%%
%%%%%%%%%%%%%%%%%%%%%%%%%%%%%%%%%%%%%%%%%%%%

Second generation interferometers are also expected to form a network of ground based detectors, as soon as 
Advanced Virgo and KAGRA  join the Hanford and Livingston LIGO sites. A collection of 
$n$ independent interferometers will roughly reduce the error by a coefficient $1/\sqrt{n}$. Looking at Fig.~\ref{fig:errors2} 
this factor would translate the network measurements at the same level of LIGO A+.
 
All the results presented so far are derived assuming a shift of the ECO's effective surface equal to $\delta=10^{-10}$, 
which (together with the mass) determines the two time-delay factors of our template. To test alternative scenarios, 
we have considered different configurations by varying $\delta$ to $10^{-20}$ and $10^{-30}$, without finding significant 
deviations from the data shown in Figs.~\ref{fig:errors1}-\ref{fig:errors2}. The parameters being mostly affected, $f_1$ and $\beta_1$, 
lead to changes $\lesssim 9\%$ and $\lesssim 2\%$, respectively, while for the other coefficients we observe variations 
below $1\%$. The values of $\epsilon_{t_\tn{echo}}$ 
and  $\epsilon_{\Delta t}$ do actually change, although the corresponding absolute errors remain constant. 
This means that the uncertainties for the new values of $\delta$ can simply be obtained from the previous results, 
by rescaling
\begin{equation}\label{scaledt}
\epsilon_{t_\tn{echo}}\Big\vert_{\delta=10^{-20}}=\epsilon_{t_\tn{echo}}\Big\vert_{\delta=10^{-10}}\frac{t_\tn{echo}(\delta=10^{-10})}{t_\tn{echo}(\delta=10^{-20})}\ ,
\end{equation}
and the same for $\Delta t$.

%%%%%%%%%%%%%%%%%%%%%%%%%%%%%%%%%%%%%%%%%%%%%%%%%%%%%%%
\subsection{echoIIa-b}
%%%%%%%%%%%%%%%%%%%%%%%%%%%%%%%%%%%%%%%%%%%%%%%%%%%%%%%

The \texttt{echoIIa} model introduces two extra parameters: (i) a second frequency within the spectrum, which leads to a 
beatlike interference with the first component, and (ii) a phase offset $\phi$ between the two echo modes. These extra parameters 
further enlarge the space of configurations to a $10\times10$ Fisher matrix.  This extension does not alter the 
estimate of $\bar{f}$ and $\bar{\tau}$, whose errors remain unchanged compared to the values obtained for the previous 
template. Moreover, as already described for \texttt{echoI}, all the results are nearly degenerate with respect to $t_\tn{echo}$, 
as changes on this parameter do not lead to sensible variations of the errors. For this reason we will only focus on the 
dependence of the echo's errors on $\beta_1$.

The parameter estimation of this toy model template shows that the results are strongly affected by the choice of $\phi$. 
In particular the error distribution finds a minimum when the echo modes are out of phase with $\phi=-\pi/2$, while 
it is maximum when the two components are on phase, i.e., $\phi=0$.
%%%%%%%%%%%%%%%%%%%%%%%%%%%%%%%
\begin{figure*}[ht]
\centering
\includegraphics[width=5cm]{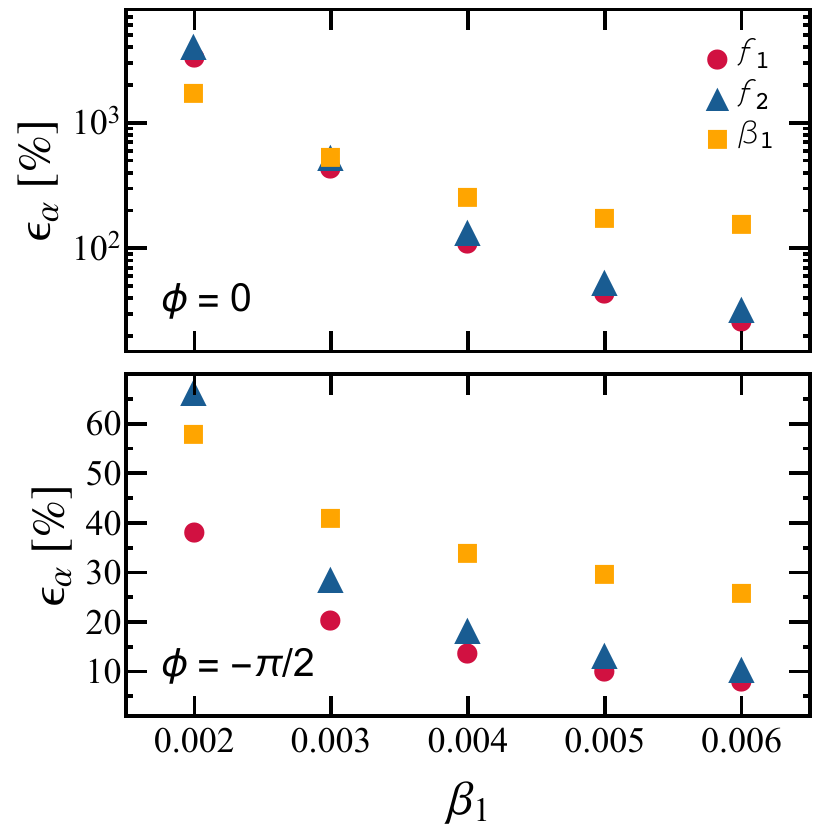}
\includegraphics[width=5.15cm]{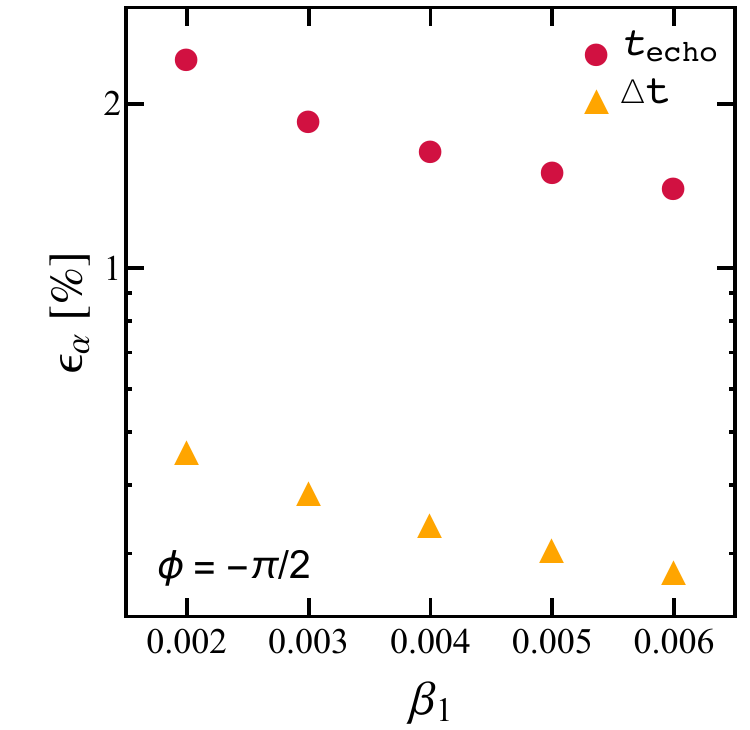}
\includegraphics[width=5.15cm]{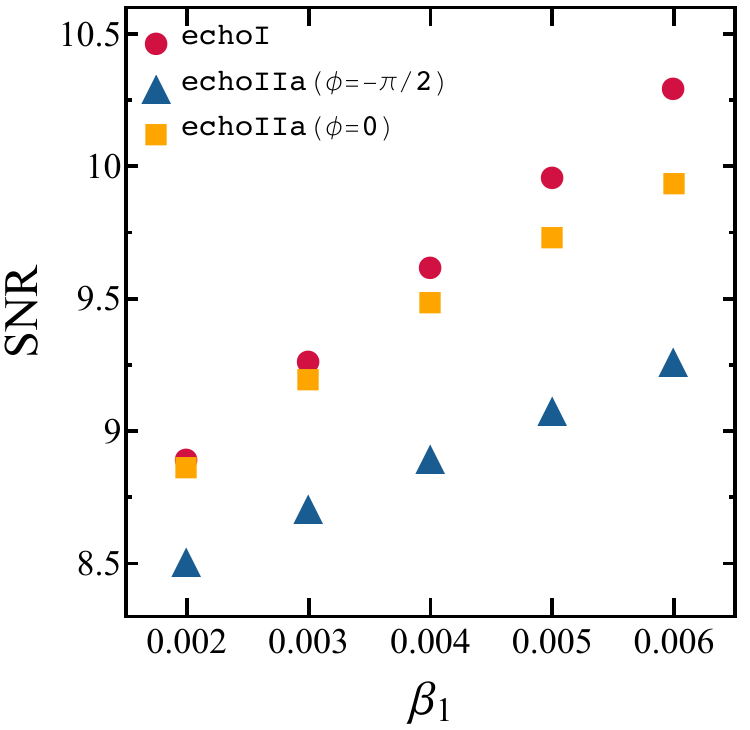}
\caption{(Left and center panels) Same as Fig.~\ref{fig:errors1} but for the parameters of the \texttt{echoIIa} model with phase 
shift $\phi=0$ and $\phi=-\pi/2$. (Right panel) Comparison between the SNR of \texttt{echoI} and \texttt{echoIIa} as a function of 
$\beta_1$. All the results refer to Advanced LIGO, assuming $t_\tn{echo}\approx \Delta t$, with $\delta=10^{-10}$.}
\label{fig:errors3}
\end{figure*}
%%%%%%%%%%%%%%%%%%%%%%%%%%%%%%%%%%%%%%%%%%%%
This effect is particularly relevant for $(f_1,f_2,\beta_1)$, as shown in the left panel of Fig.~\ref{fig:errors3} in which we 
draw the corresponding relative uncertainties computed for LIGO, assuming $\delta=10^{-10}$.
As already seen for \texttt{echoI}, all the errors decrease with the growth of the shape factor, up to our best model 
with $\beta_1=0.006$. The two frequencies yield almost the same accuracy for $\phi=0$, while for out-of-phase modes 
the errors on $f_2$ are in general larger and converge to $\epsilon_{f_1}$ for $\beta_1\gtrsim0.004$ only. Figure \ref{fig:errors3} 
also shows that our ability to measure the width of the Gaussian function strongly depends on the phase of the echoes, 
as for $\phi=0$ all of the configurations lead to errors above an upper bound $\epsilon_{\beta_1}=1$. This picture 
changes dramatically if $\phi=-\pi/2$, for which the uncertainties on this parameter is of the same order of magnitude as 
$(f_1,f_2)$, and smaller than  $50\%$ for $\beta_1>0.002$.
Variations of $\phi$ are subordinate on $t_\tn{echo}$ and $\Delta t$, for which we observe small deviations between 
the two cases. The center panel of Fig.~\ref{fig:errors3} shows that even for this template the two parameters 
provide the best measurements, at the level of percent and below. 

\begin{figure}[ht]
\centering
\includegraphics[width=6cm]{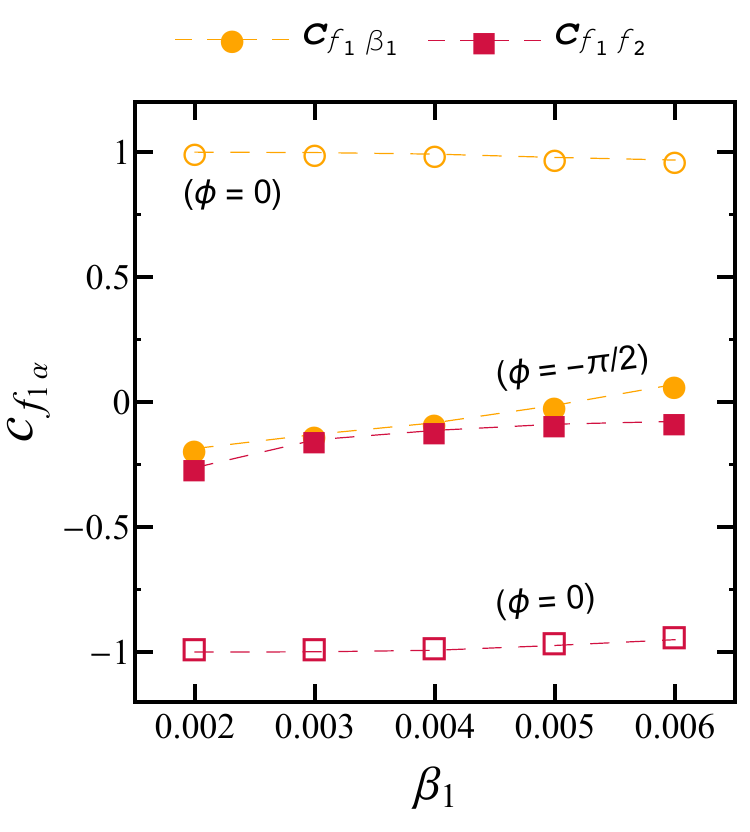}
\caption{Correlation coefficients for the \texttt{echoIIa} template between $f_1$ and $(\beta_1,f_2)$, assuming $\phi=0$ (empty dots) 
and $\phi=-\pi/2$ (empty dots), for LIGO with $\delta=10^{-10}$ and $t_\tn{echo}\approx \Delta t$.}
\label{fig:correlation}
\end{figure}

As expected, even for the most optimistic scenario ($\phi=-\pi/2$), the results obtained for this model are 
in general worse than those derived for the \texttt{echoI} (cf. Fig.~\ref{fig:errors1}). This change is partially due to 
the larger number of parameters which, for a given configuration and detector sensitivity, dilutes the amount of information 
contained within the waveform. In this regard, it is also interesting to compare how the specific form of the waveform 
may influence the SNR of the signal (and the degeneracies between the parameters). The right panel of Fig.~\ref{fig:errors3}
shows indeed how this quantity changes as a function of $\beta_1$ for \texttt{echoI} and \texttt{echoIIa} (and two values of 
$\phi$). The picture leads to some interesting conclusions. First, we observe that for all cases considered 
the second parametrization yields lower SNR. Moreover, the overall growth is softer, with an increase of  $\sim9\%$-$12\%$ 
(depending on the value of $\phi$) compared to a change of $\sim16\%$ for the first template. 
More significantly, the values of $\rho$ for $\phi=-\pi/2$ are always smaller than those for $\phi=0$, which is a rather counterintuitive result, as the errors scale in the opposite direction. In this case a major role is played by the 
degeneracy between the parameters, as shown in Fig.~\ref{fig:correlation} where the correlation 
coefficients $c_{\alpha\beta}=\Sigma_{\alpha\beta}/(\sigma_\alpha\sigma_\beta)$ between $f_1$ and $(\beta_1,f_2)$ are plotted. 
For $\phi=0$ the two components of the echoes \eqref{heco2a} are described exactly by the same functional form, and 
all the variables are extremely correlated, as for the two frequencies for which $c_{f_{1}f_2}\simeq-1$. 
Note that in the limit $f_1\rightarrow f_2$ we would have $100\%$ degeneracy. 
Conversely, for out-of-phase modes with $\phi=-\pi/2$ we have a maximum break of such degeneracy that allows one to 
set tighter constraints on the parameters. Moreover, the values of $c_{f_1\beta}$ for 
$\beta=\{t_\tn{echo},\Delta t\}$ are always close to zero for any choice of $\phi$, which is in line with the results previously 
described.

%%%%%%%%%%%%%%%%%%%%%%%%%%%%%%%
%%%%%%%%%%%%%%%%%%%%%%%%%%%%%%%%%%%%%%%%%%%%

Finally, unlike the \texttt{echoI}, the second template is more sensible to different values of $\delta$. 
Comparing the results obtained for $\delta=10^{-30}$ and  $\delta=10^{-10}$, assuming the optimal 
case $\beta_1=0.006$ and $\phi=-\pi/2$, we find that the absolute errors of $\{f_1,\beta_1,f_2,t_\tn{echo},\Delta t\}$ vary 
approximately as $\simeq\{23,-12,-8,-6,-41\}\%$, and therefore a scaling such as that 
given by Eq.~\eqref{scaledt} is no longer valid. These differences grow dramatically for $\phi=0$.\\

\begin{figure}[ht]
\centering
\includegraphics[width=9cm]{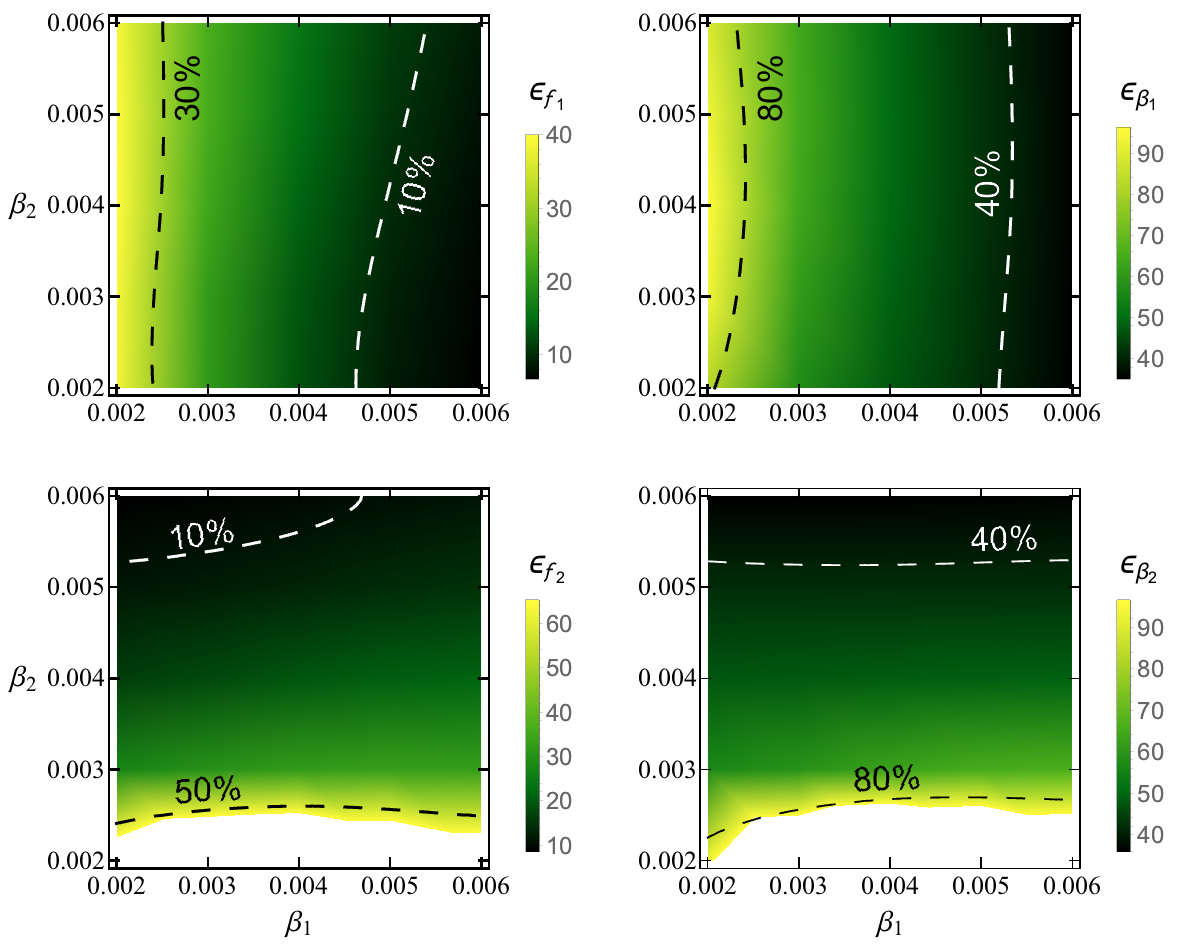}
\caption{Same as Fig.~\ref{fig:contours} but for the parameters of \texttt{echoIIb} in the $\beta_1\times\beta_2$ plane. 
The phase offset is fixed to $\phi=-\pi/2$.}
\label{fig:contours2}
\end{figure}

As the last step we analyze the output of the \texttt{echoIIb} model, which improves the former description by 
adding another shape factor ($\beta_2$) to the second component of the pulses, specified by the frequency $f_2$. 
For the sake of simplicity, in this case we will fix $\delta =10^{-10}$ and $t_\tn{echo}=\Delta t$.
Then, we span the possible configurations within the 
$\beta_1\times \beta_2$ parameter space, also assuming the two phase shifts considered before, i.e., $\phi=0$ and 
$\phi=-\pi/2$. Figure \ref{fig:contours2} shows the numerical results obtained for the latter, assuming the LIGO detector. 

From the top panels we observe that the relative errors on $(f_1,\beta_1)$ are nearly degenerate with respect to the Gaussian 
width of the second component. The opposite occurs if we look at the behavior of $\epsilon_{f_2}$ and $\epsilon_{\beta_{2}}$ 
in the bottom plots. This feature is mainly due to the specific form of the template, such that the diagonal components of 
the Fisher matrix for the first mode is independent of the second one and vice versa. In both cases, however, a sweet spot 
exists for larger values of the shape factors that yield the best results. This clearly confirms the trend observed for 
the \texttt{echoI} and \texttt{echoIIa} templates. 
Note also that the parameter's accuracy of both modes is comparable. Only a few configurations, clustered around $\beta_2\simeq0.002$, 
lead to errors on $f_2$ and $\beta_2$ exceeding the upper bound $\epsilon_\alpha=1$. The relative uncertainties on  
the time shifts (not shown in the figure) are in agreement with the results obtained for the previous waveforms,  
with $\epsilon_\tn{echo}\lesssim3\%$ and $\Delta t\lesssim 1\%$ for all the points in the $\beta_1\times\beta_2$ plane.

%%%%%%%%%%%%%%%%%%%%%%%%%%%%%%%
%%%%%%%%%%%%%%%%%%%%%%%%%%%%%%%%%%%%%%%%%%%%

\begin{figure}[ht]
\centering
\includegraphics[width=4.25cm]{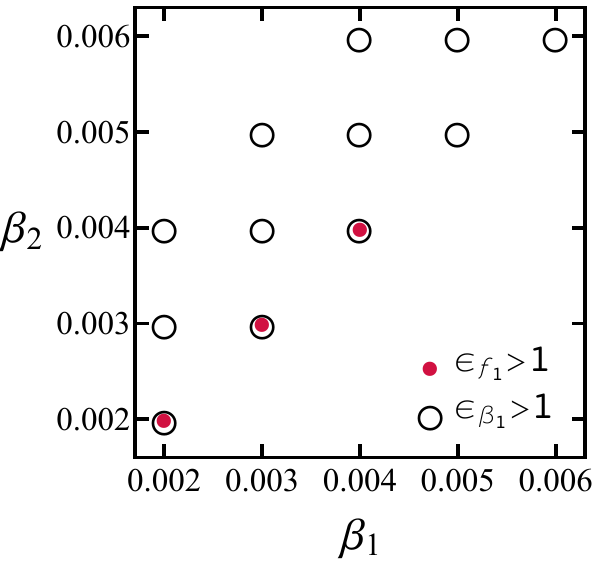}
\includegraphics[width=4.25cm]{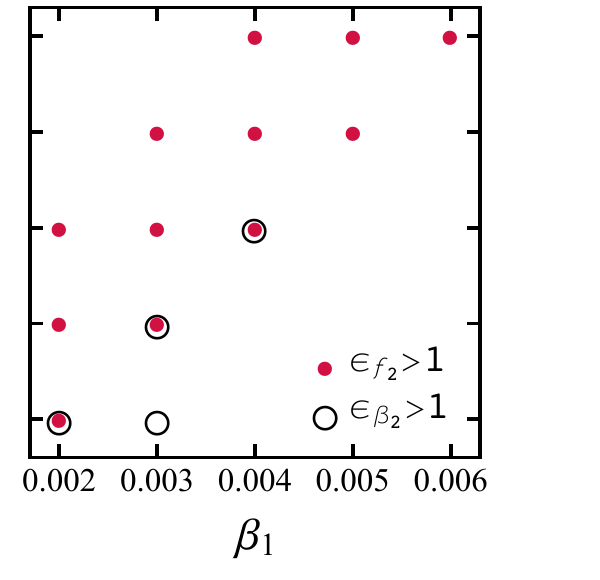}
\caption{Each dot specifies a configurations in the $\beta_1\times\beta_2$ plane with $\phi=0$ for which the relative error 
$\epsilon_{\alpha}>1$.}
\label{fig:upperbound}
\end{figure}

A phase shift $\phi=0$ between the echo's components would, again, reduce our ability to detect frequencies and 
shape factors. This is shown in Fig.~\ref{fig:upperbound}, in which each point identifies a specific configuration 
for which the relative error of a certain parameter is larger than $1$, i.e., for which its measurement is strictly 
compatible with zero.

%%%%%%%%%%%%%%%%%%%%%%%%%%%%%%%
%%%%%%%%%%%%%%%%%%%%%%%%%%%%%%%%%%%%%%%%%%%%

\begin{figure}[ht]
\centering
\includegraphics[width=7cm]{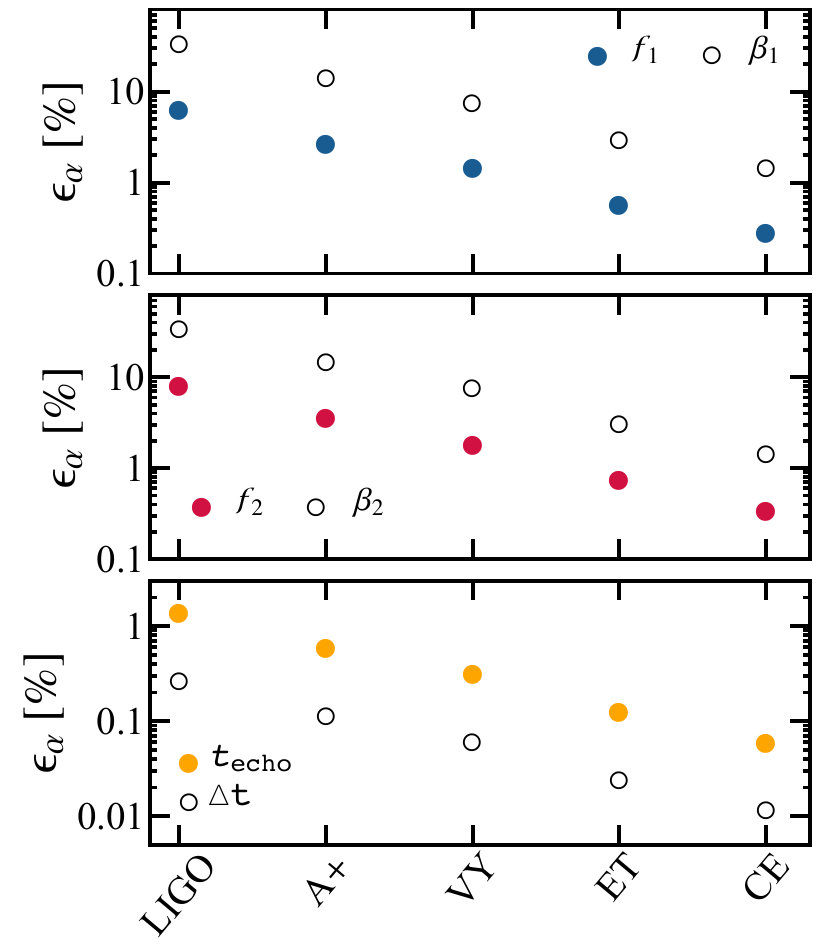}
\caption{Relative errors of the  \texttt{echoIIb} model for the best case scenario with 
$\phi=-\pi/2$, computed for different interferometers.}
\label{fig:errors4}
\end{figure}

This picture rapidly improves for future detectors as demonstrated in Fig.~\ref{fig:errors4}, in which we plot 
the errors corresponding to the best configurations, with $\phi=-\pi/2$. Note that the most accurate results 
occur when the difference between the two Gaussian widths is maximum, i.e., when $(\beta_1,\beta_2)=(0.006,0.002)$
for $\epsilon_{f_1}$ and $\epsilon_{\beta_1}$, and when $(\beta_1,\beta_2)=(0.002,0.006)$ for $\epsilon_{f_2}$ and $\epsilon_{\beta_2}$.
The picture shows, for example, that LIGO A+ (which we remind the reader roughly corresponds to a network of current detectors) would 
already constrain frequencies and shape 
factors with a relative accuracy around $\ll10\%$. A third generation detector like the Einstein Telescope would be required to reduce these errors below $1\%$.

%%%%%%%%%%%%%%%%%%%%%%%%%%%%%%%
%%%%%%%%%%%%%%%%%%%%%%%%%%%%%%%%%%%%%%%%%%%%

%%%%%%%%%%%%%%%%%%%%%%%%%%%%%%%%%%%%%%%%%%%%%%%%%%%%%%%
\section{Conclusions}\label{Sec:conc}
%%%%%%%%%%%%%%%%%%%%%%%%%%%%%%%%%%%%%%%%%%%%%%%%%%%%%%%

Gravitational wave astronomy is establishing itself as a new field of research capable of gaining insights on 
a genuine strong field gravity regime, and answering open questions of fundamental physics. A key example is 
given by the possible existence of horizonless exotic objects whose compactness approaches the BH limit. Such 
ECOs may form in nature and merge within the Hubble time, mimicking  the last stage of coalescence of two 
ordinary BHs \cite{Cardoso:2017cfl}. In the postmerger phase, these objects would 
produce characteristic echoes, which at late time differ from the standard QNM spectrum, and can in principle 
be detected by laser interferometers. Although current GW data seem to show no statistical evidence 
of possible deviations from the standard BH picture, it is expected that signals with larger SNR will provide 
new precious information. For this purpose it is mandatory to construct GW templates as accurately as possible, 
which allow one to capture the dominant features of the process. Recent efforts have already been done to 
build fully analytical waveforms for data analysis strategies \cite{Nakano:2017fvh,2017arXiv170606155M}. 
In this paper we pursue a complementary path, trying to address for the first time the level of accuracy with which 
current/near future interferometers will be able to detect the echo's parameters. 
To this aim we have adopted phenomenological templates depending on a relatively small set of coefficients, which try 
to mimic the expected {\it true} signal with an increasing degree of realism.

The numerical results obtained for all the considered models seem to suggest that even current detectors, 
at design sensitivity, can provide reliable estimates of all the parameters.  
Moreover, the analysis performed for the templates highlights some common properties, which can be described as follows:
\begin{itemize}
\item The SNR and the errors of realistic echo signals are expected to saturate after a certain number of 
repeated reflections, as the amplitude of each pulse decreases in time.
\item The uncertainties on the template's parameters are mostly affected by the width of the function that 
shapes the echoes. In particular, larger values of this factor always lead to an increase of the overall SNR and to 
a reduction of the errors. 
\item As far as multiple frequencies are considered, the phase offset between different components of the echoes 
plays a crucial role, and it strongly affects the degeneracy between the parameters. Modes out of phase (in phase) 
lead to minimum (maximum) errors. 
\item Best case scenarios for all the models show that the frequencies and the shape factors of the echoes can always 
be measured with an accuracy smaller than $100\%$. A network of advanced detectors,  composed 
of the two LIGO, Virgo and KAGRA, would reduce these values around $10\%$. Third  generation interferometers, 
such as the Einstein Telescope, are required to measure the same quantities at the level of percent.
\item The parameters that characterize the time delay between the BH QNM component and the subsequent echoes 
are measured with exquisite accuracy, with relative errors $\lesssim3\%$ with Advanced LIGO already. Moreover
changes in $t_\tn{echo}$ and $\Delta t$ seem to slightly affect the other parameters of the waveform. 
\item Complex templates, in which multiple frequencies may interfere to produce the echoes, 
are more sensible to variations of the parameter $\delta$ which controls the shift between the ECO's surface with 
respect to a Schwarzschild BH horizon. 
\end{itemize}

A summary of the results for the most complex model can be found in Figs.~\ref{fig:contours2}-\ref{fig:errors4}. 

The data analysis developed in this paper may be considered as a proof of principle for future 
developments, and it suffers from two main limitations. The first obvious 
drawback is given by the lack of a semianalytical template able to fully characterize the GW emission of perturbed 
ECOs. The phenomenological models used here represent a first step in this direction, which provide a reliable 
description of the full picture, being still based on a limited numbers of parameters. 
Note that, unlike standard QNMs, which are solely determined by the BH mass and spin, the echo structure 
is intrinsically more complex as follows: (i) the trapped modes spectrum crucially depends on the specific ECO considered, 
and (ii) the shape of the echoes can be affected by the specific form of the perturbation. A second source of uncertainty, 
connected to the previous problem, relies on the unique identification of the echo's amplitudes. Although the assumptions employed in Sec.~\ref{Sec:templates} 
are physically motivated by well known results \cite{Kokkotas:1995av}, our conclusions still depend on the 
 relative strength of the pulses and can be considered as an optimistic scenario.

Improvements of the previous points can pursue various directions. 
Our current efforts are particularly devoted to investigate in detail the following aspects: 
(i) construct more refined models that approach realistic ultracompact objects with larger accuracy, 
possibly taking into account the interference of multiple trapped modes;
(ii) use a fully Bayesian analysis to perform model selection and assess the 
ability of the ground based interferometer to distinguish between standard BH and echolike signals;
and (iii) employ realistic errors to reconstruct the ECO's scattering potential by measurements of the 
trapped modes, as done in \cite{paper2}. These extensions are already under investigation. 

%%%%%%%%%%%%%%%%%%%%%%%%%%%%%%%%%%%%%%%%%%%%%%%
\noindent{\em Acknowledgements.---}
S.V. is grateful for the financial support of the Baden-W\"urttemberg Foundation. 

%%%%%%%%%%%%%%%%%%%%%%%%%%%%%%%%%%%%%%%%%%%%%%%

\bibliography{literatur1}

\end{document}